\documentclass{ws-p8-50x6-00}
\def\lsimeq{\,\,\raise0.14m\hbox{$<$}\kern-0.76em\lower0.28em\hbox
{$\sim$}\,\,}

\begin{document}
\title{INTEGRAL and Nuclear Astrophysics}
\author{Jacques Paul, Michel Cass\'e}
\address{SAp/DAPNIA/DSM/CEA, Orme des Merisiers,
 91 191 Gif/Yvette CEDEX, France and \\
 Institut d'Astrophysique de Paris, CNRS, 98 bis bd Arago, 75014 Paris, France}
\author{Elisabeth Vangioni-Flam}
\address{Institut d'Astrophysique de Paris, CNRS, 98 bis bd Arago, 75014 Paris, France}
\maketitle\abstracts{We briefly review the fundamentals of nuclear gamma-ray line astronomy 
(radioactive astronomy), focusing on its role to decipher the intimate physics of supernovae,
 either immediatly (via $^{56}Co)$ or after a time delay (via $^{44}Ti$). 
All kinds of supernovae can be
 in principle tested through their radioactivities and their associated gamma-ray lines.
 Dedicated to the spectroscopy and imaging of celestial sources in the 15 keV to 10 MeV band,
 the ESA scientific observatory INTEGRAL will
 open a golden age of nuclear astrophysics in Europe.}

\section{Why a new gamma-ray astronomy mission?}

Gamma-ray astronomy explores the most energetic phenomena
 that occur in 
nature  
and addresses some of the most fundamental problems in    
astrophysics. It  
embraces a great variety of gamma-ray continuum
 and gamma-ray line  production  
processes:  
nuclear excitation, radioactivity, positron annihilation and Compton     
scattering; and  
an even a greater diversity of astrophysical objects and phenomena:     
nucleosynthesis,  
nova and supernova explosions, the interstellar medium, cosmic-ray     
interactions and  
sources, neutron stars, black holes, gamma-ray bursts, active galactic     
nuclei and the  
cosmic gamma-ray background. Not only do gamma rays allow us to see     
deeper into  
these objects, but the bulk of the power radiated by them is often at     
gamma-ray  
energies.

In the low-energy gamma-ray band, line-forming processes such as nuclear     
excitation, radioactivity, positron annihilation, cyclotron emission and     
absorption  
become important, and when used as astrophysical tools. Unique astrophysical information is     
contained in the  
spectral shift, line width, and line profiles. Detailed studies of these     
processes require  
the resolving power of a germanium spectrometer. Lower-resolution     
spectrometers  
(e.g. SIGMA, OSSE, COMPTEL) did not have sufficient energy resolution to     
permit  
a study of the parameters of these lines. The last high-resolution space     
instrument,  
that on HEAO-3 in 1979-80, was 100 times less sensitive than 
required to tackle  
the scientific subjects outlined below. Solid observational and     
theoretical  
grounds already exist for predicting detectable line emission from such     
varied  
celestial objects as the Central region of the Galaxy, the interstellar     
medium, compact  
objects, novae and supernovae and a variety of active galactic nuclei.

\section{Nuclear astrophysics and gamma-ray astronomy}

\subsection{Generalities}

The observability of gamma-ray lines is conditionned by three factors:
 
a. Thermonuclear synthesis in stellar furnaces of fresh radioisotopes in 
significant abundance (N).

b. Quick removal (before decay) from the dense medium of birth to avoid 
burning and allow transfer to low density regions, transparent to gamma rays.
 
c. Lifetime sufficiently  short ( radioactive constant, $\lambda$, high enough) in order 
to get measurable activity ($\lambda$N), i.e. enough photons  to be detected at 
galactic/cosmic distances. 

The list of candidates consequently is rather brief; it includes principally $^{56}Co$, 
 $^{44}Ti$ and $^{26}Al$ (the lifetimes are respectively 112 days, 87 years, $10^6$ years).

The observation of nuclides through the gamma-ray lines emitted in the course 
of their natural decay in the circumstellar/interstellar media reveals the 
recent nucleosynthesis activities on time scales comensurate to their lifetime. 
According to the species involved, one can probe individual events, like 
supernovae and novae or collective effects of these objects at galactic scale.
The two limiting cases are the following:

i) if the lifetime is short($^{56}Co$) compared to the frequency of the relevant event  (SN 
explosion in this case) one has the possibility of observing 
individual events.

ii) if the  lifetime is long compared to the frequency of fertile events ($^{26}Al$) one should observe the 
collective effect of many events under the form of a diffuse galactic emission. 
For lifetimes of the order of the recurrence time (rate) , i.e. $^{22}Na$ and
 $^{44}Ti$, the case is 
intermediate. One should then observe diffuse extended sources.

The line profiles offer a wealth of information on the detailed physics/ 
dynamics of the astrophysical sources of unstable isotopes \cite{di} \cite{K}. 
Gamma ray lines are indeed ideal diagnostics of nucleosynthesis. Stringent constraints 
on stellar and supernova yields can be set by specific line intensity or even 
upper limits on these. 
Gamma-ray line astronomy, in principle should allow to test both quiescent 
(hydrostatic) nucleosynthesis (in AGB and Wolf- Rayet stars) and explosive 
nucleosynthesis (novae and supernovae). We focus here on supernovae (for 
the other objects see J. Knodlseder and M. Hernanz, this conference). We are 
concerned here by the high potential of these exploding objects in both the synthesis 
and acceleration of nuclei.

\subsection{Gamma-ray lines and supernova physics}

A highligt of gamma-ray line astronomy has been the observation of 
signatures of the radioactive decay (weak interaction) of $^{56,57}Co$ isotopes, 
implied in the $^{56,57}Ni$ chain ending at $^{56,57}Fe$ from the supernova SN1987A in the Large 
Magellanic Cloud.
Radioactive decay proceeds, in all cases considered, through the weak 
interaction ($\beta^+$ decay or electron capture) transforming inside a nucleus a 
proton into neutron. Indeed the synthesized species, due to the physics 
involved, are too proton rich to be stable. This proton richness is itself related 
to the fact that most of the reactions that built complex nuclei are induced by protons 
and alphas. This is especially true for explosive nucleosynthesis in supernovae, 
where the symetrical $^{44}Ti$ (22 protons and 22 neutrons) and $^{56}Ni$ (28 protons, 
28 neutrons) are formed by successive alpha captures on $^{28}Si$.
 
All kinds of supernovae can be tested (Ia,b,c and II) both through their 
radioactivity and also their (collective) particle acceleration effect.

i) individual SNII: There is very little chance to capture the cobalt-iron 
gamma-ray line emission of a new core collapse supernova during the lifetime
 of the present generation
 of gamma-ray satellite. However young 
supernova remnants could show up through the $^{44}Ti$ lines
and through superbubbles excavated in the interstellar medium by the combined effect 
of strong stellar winds of massive stars and explosions.
 Due to its lifetime, of the order of a century,  $^{44}Ti$ is well 
suited to discover young SNR  hidden in the galactic dust. $^{44}Ti$ decay lines 
from recent  SNR (Cas A and Vela junior) without optical counterpart have already
been detected. Thus we have at least two cases of hidden supernovae. How many others will 
 the INTEGRAL mission discover? The estimates are rather uncertain \cite{TH}.
The detailed exploration of  the known sources, associated to X ray data from 
Chandra and XMM, should allow to derive the distribution of $^{44}Ti$ and its mass 
(taking into account the fact that its lifetime against electron capture is possibly 
modified by very high state of ionisation of the emitting regions), setting 
important constraints on the physics of the explosion.
  
ii) collective effects of SNII in Superbubbles: Non thermal nucleosynthesis of 
Lithium-Beryllium-Boron (see M. Cass\'e, this conference) could 
also be accompanied by specific (rather wide) gamma ray lines, specially those 
of C*, O* and the LiBe feature arising from the alpha + alpha reaction \cite{PA} \cite{TA}. 
Indeed, superbubbles seem to be the most 
favourable sites of production of non thermal gamma-ray lines. 

iii) SNIa: A direct observation of the gamma  ray lines associated to $^{56}Co$ 
from a SNIa (closer than 15 Mpc) would be a 
very happy event. It could serve to clarify the thorny physics of these objects, 
which serve, on the other hand, as distance indicators for cosmology.

\section{ The INTEGRAL mission}

\subsection{Overall presentation}

Dedicated to the spectroscopy and imaging of celestial sources in the 15     
keV to  
10 MeV band, the ESA (European Space Agency) scientific mission INTEGRAL     
(International Gamma-Ray Astrophysics Laboratory) will address the     
scientific  
objectives defined above through the simultaneous use of two main gamma-ray     
instruments, the  
high resolution spectrometer SPI and the fine imaging telescope IBIS,     
with  
concurrent source monitoring in the X-ray and visible bands. High     
resolution  
spectroscopy with fine imaging and accurate positioning of celestial     
sources over the  
entire energy range are mandatory to reach the scientific goals of the     
mission. Fine  
spectroscopy over the entire energy range will permit spectral features     
to be uniquely  
identified and line profiles to be determined for physical studies of     
the source region.  
Fine imaging capability within a large field of view will permit the     
accurate location  
and hence identification of the gamma-ray emitting objects with     
counterpart at other  
wavelengths. It will also enable extended regions to be distinguished     
from point  
sources and provide considerable serendipitous science. Explicit     
references on  
INTEGRAL science can be found in the proceedings of past INTEGRAL     
Workshops  
held in les Diablerets \cite{jac1}, Saint Malo \cite{jac2}, Taormina \cite{jac3}, and Alicante \cite{jac4}.

The INTEGRAL spacecraft (Figure 1) consists of a service module,     
commonly  
designed with the service module of the ESA XMM-Newton mission    
containing all  
spacecraft subsystem and a payload module containing the scientific     
instruments.  
INTEGRAL, with a payload mass of approximatly 2000 kg and a total launch mass     
of about 4000 kg  
will be launched in April 2002 by a Russian PROTON launcher into a     
highly  
eccentric 72-hour orbit (initial perigee height 10 000 km, initial     
apogee height  
153 000 km). The particle induced background affects the performance of     
high-energy detectors, and scientific observations will therefore be carried     
out while the  
spacecraft is above an altitude of nominally 40 000 km, implying that     
90$\%$ of the  
time spent on the orbit can be used for scientific observations.     
 However, when taking into account several in-orbit
 activities (as e.g. slew and instrument calibration), the average observation
 efficiency becomes about 85 $\%$ per year. The spacecraft employs fixed 
 solar arrays: this means, that the target pointing of the spacecraft
 (at any point in time) will remain outside an avoidance
 cone around the sun and anti-sun. This leads to a minimum angle
 between any celestial source and the sun/anti-sun of 50 degrees
  during the nominal mission life.

 INTEGRAL will be an observatory-type mission with a nominal
 lifetime of 2 years, an extension up to 5 years is technically possible. 
 Most of the observing time (65$\%$ during year 1, 70 $\%$ year 2 and 
 75 $\%$ after) will be awarded to the scientific community at large
 as the General Program. Typical observations will last up two weeks.
 Proposals following a standard
 AO process, will be selected on their scientific merit only by
 a single Time Allocation Committee.
 The first call for observation proposals, released in November 2000, has
 been closed on February 16, 2001. The remaining fraction of the observing
 time will be reserved, as guaranteed time, for the INTEGRAL Science Working
 Team for its contribution to the program. This fraction, the Core Program,
 will
 be devoted a Galactic plane survey, a deep exposure of the central radian
 of the Galaxy and pointed observations of selected regions 
(such as the Vela region) and targets of opportunity (TOO).
 In accordance with ESA's policy on data rights, all scientific 
 data will be made available to the scientific community at large 
 one year after they have been released to the observer. For more information see the INTEGRAL  World Wide Web pages\cite{jac5}. 

The INTEGRAL Science Data Center (ISDC) located in Versoix, Switzerland, close
 to the Geneva Observatory, will be the center in which the INTEGRAL payload 
 telemetry data will be processed to a level at which all users can pursue the
 scientific interpretation of the data. The data will be corrected for instrumental
 signature and 
some standard scientific processing and analysis will be performed. The     
ISDC is the  
place where the archive and derived products will be built and made     
accessible to the  
world wide astronomical community. The ISDC will routinely monitor the     
instrument  
science performance and conduct a quick-look science analysis. Most of     
the TOO  
showing up during the lifetime of INTEGRAL will be detected at the ISDC     
during the  
routine scrutiny of the data. Scientific data products obtained by     
standard analysis  
tools will be distributed to the observers and archived for later use by     
the science  
community.

\begin{figure}[h]
\vspace*{-20ex}

\epsfxsize  25pc 
\epsfbox{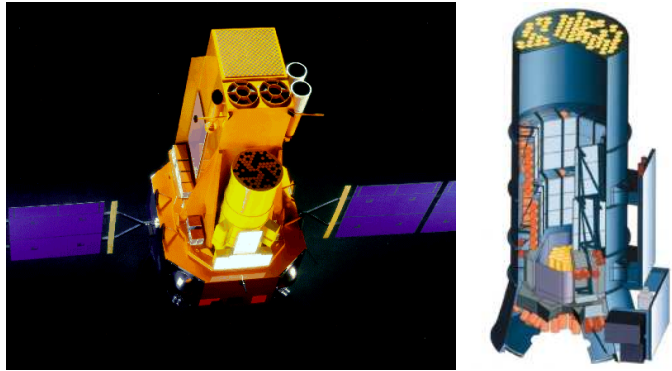} 
\vspace*{-30ex}

\caption{ (Left) The INTEGRAL spacecraft. The cylindrically shaped SPI is next     
to the  
larger rectangular payload module (PLM) structure housing the IBIS and     
JEM-X  
detectors inside. The top of the PLM carries the coded mask for IBIS     
(squared) and  
the two coded masks for the two JEM-X detectors. The OMC is located at     
the left top  
of the PLM. (Right) Schematic view of the spectrometer SPI.}
\end{figure}

\subsection{The spectrometer SPI}

One of the two main INTEGRAL instruments, the spectrometer SPI (Figure     
1), has  
been designed to perform spectral analysis of localized and extended     
regions with  
performances at 1 MeV such as an energy resolution $E/\Delta E \sim 500$     
and a $3 \sigma$  
line sensitivity of $5\ 10^{-6}$ photons ${\rm s}^{-1}\ {\rm cm}^{-2}$ ($10^6$ s exposure).
 The     
instrument features an  
array of 19 hexagonal high purity germanium detectors cooled by an     
active device to  
an operating temperature of 85 K. An hexagonal coded aperture mask is     
located 1.7  
m above the detection plane in order to image large regions of the sky     
(fully coded  
field of view $\sim$ 16 degrees) with an angular resolution of 2 degrees.
 In     
order to reduce background  
radiation, the detector assembly is shielded by a veto system which     
extends around  
the bottom and side of the detector almost completely up to the coded     
mask. A plastic  
veto is provided below the mask to further reduce the 511 keV     
background. The  
spectrometer SPI is being developed under prime contractor     
responsibility of CNES  
(the French Space Agency) by a consortium of institutes in France (CESR     
Toulouse,  
CEA Saclay), Germany (MPE Garching), Italy (IFCTR Milano), Spain (U     
Valencia),  
Belgium (U Louvain), UK (U Birmingham), USA (UC San Diego, LBL Berkeley,     
NASA/GSFC Greenbelt). Principal Investigators are G. Vedrenne (CESR     
Toulouse),  
and V. Schoenfelder (MPE Garching).

\section{Conclusion}

With the launch in the near future of the INTEGRAL spacecraft, a golden     
age of  
nuclear astrophysics will open in Europe, at the point of convergence of     
nuclear  
physics and astrophysics. Time is ripe to join this effort, especially     
having in mind  
the fact that INTEGRAL is an observatory-type mission with most of the     
total  
observing time being awarded as the general program to the scientific     
community at  
large.


\begin{thebibliography}{99}
\bibitem {di} R. Diehl, in "Astronomy with Radioactivities", 
Workshop Proceedings, Ringberg, sept 1999, MPE Report,  {274} {3} {1999}
\bibitem{K} J. Knodlseder, in "The Interplay Between Massive Stars and the ISM" 
parallel session of JENAM99, sept 1999, Toulouse,
 Edts D. Schaerer and R. Gonzalez-Delgado, astro-ph/9912131
\bibitem{TH} L.S. The, in "Astronomy with Radioactivities", 
Workshop Proceedings, Ringberg, sept 1999, MPE Report,  274, p.77, 1999
\bibitem{PA} E. Parizot, M. Cass\'e and E. Vangioni-Flam \Journal {\it A\&A}
 {328} {107} {1997}
\bibitem{TA} V. Tatischeff et al, in the 4th INTEGRAL Workshop
 "Exploring the Gamma-Ray Universe", Alicante, sept. 2000, in press

\bibitem {jac1} Proc. 1st INTEGRAL Workshop, The Multiwavelength Approach to     
Gamma-Ray  
    Astronomy, ApJS 92, No. 2, 1994
\bibitem {jac2} Proc. 2nd INTEGRAL Workshop, The Transparent Universe, Eds: C.     
Winkler,  
T.  J.-L. Courvoisier, Ph. Durouchoux, ESA SP-382, 1997
\bibitem {jac3} Proc. 3rd INTEGRAL Workshop, The Extreme Universe, Astrophys. Let. \&     
Communications, 38 (Part I) \& 39 (Part II), 1999
\bibitem {jac4} Proc. 4th INTEGRAL Workshop, Exploring the Gamma-Ray Universe, 2001 in     
press
\bibitem {jac5} http://astro.estec.esa.nl/SA-general/Projects/Integral/integral.html

\end{thebibliography}
\end{document}